
%
%
%
%
%

\input amssym.def \input amssym.tex

\magnification=1200
\def\Cal{\cal} \def\roman{\rm}
\def\Ga{\Gamma}
\def\Go#1{\Ga_0(#1)}
\def\Goo#1{\Ga_0^0(#1)}
\def\Gmn#1#2{\Ga(#1,#2)}
\def\Gn#1{\Ga(#1)}
\def\Gg{\Ga_g}
\def\ref#1{[{\bf #1}]}
\def\Tr#1#2{{\roman Tr}_{#1}(#2 q^{L_0-1})}
\def\V{{\Cal V}} \def\Vg{\V_g} \def\Vh{\V_h} \def\Vgh{\V_{gh}}
\def\Vr{\V_r} \def\Va{\V_a}  \def\Vst{\V^\ast}
\def\Vf{\V_f}
\def\H{{\Cal H}} \def\Hg{\H_g} \def\Hh{\H_h} \def\Hgh{\H_{gh}}
   \def\Hst{\H^\ast}

\def\M{{\Bbb M}}
\def\gen#1{\langle #1 \rangle}
\def\om{\omega}
\def\MM{{\V}^\natural} \def\HMM{{\Cal H}^\natural}
\def\L{\Lambda} \def\Lst{\Lambda^\ast}
\def\VL{{\V}^\L} 
\def\Z{{\Bbb Z}} \def\R{{\Bbb R}}
\def\Vorb#1{\V_{\roman orb}^{\gen{#1}}}
\def\Vorbg{\Vorb{g}}  \def\Vorbh{\Vorb{h}} \def\Vorbgh{\Vorb{g,h}}
\def\VorbG{\V_{\roman orb}^G}  \def\Vorba{\Vorb{a}} \def\Vorbf{\Vorb{f}}

\def\mod#1{{\ \roman mod\ }#1}
\def\con#1{{\buildrel #1\over\sim}}
\def\m1{{-1}}
\def\P{{\Cal P}}  \def\Pg{\P_{\gen{g}}}
\def\Ph{\P_{\gen{h}}} \def\Pgh{\P_{\gen{g,h}}} \def\Pf{\P_{\gen{f}}}
\def\Pa{\P_{\gen{a}}} \def\PG{\P_G} \def\Pr{\P_{\gen{r}}}
\def\vac{\vert 0\rangle}
\def\op{operator } \def\ops{operators }
\def\auto{automorphism }
\def\autos{automorphisms }
\def\r{{\overline r}}
\def\a{{\overline a}}

\def\Co{{\roman Co}_1}
\def\Tgt{T_g(\tau)}
\def\Tft{T_f(\tau)}

\hfill DIAS-STP-94-38\break
\centerline {\bf GENERALISED MOONSHINE}
\medskip
\centerline{\bf AND}
\medskip
\centerline{{\bf ABELIAN ORBIFOLD CONSTRUCTIONS}
\footnote*{Talk presented at AMS Meeting on Moonshine, the
monster and related topics, Mt. Holyoke, June 1994}}
\vskip 1truecm
\centerline{Michael P. Tuite
\footnote\dag{EMAIL: michael.tuite@ucg.ie}
}
\centerline{Department of Mathematical Physics}
\centerline {University College, Galway, Ireland}
\smallskip
\centerline {and}
\smallskip
\centerline{Dublin Institute for Advanced Studies}
\centerline{10 Burlington Road, Dublin 4, Ireland}
\vskip 1 truecm

{\bf 0. Introduction.}
We consider the application of
Abelian  orbifold constructions
in Meromorphic Conformal Field Theory (MCFT) \ref{Go,DGM}
towards  an understanding of various aspects of
Monstrous Moonshine \ref{CN} and Generalised Moonshine \ref{N}.
We review  some of the basic concepts in MCFT and
Abelian orbifold constructions \ref{DHVW} of MCFTs and
summarise some of the relevant
physics lore surrounding such constructions
including aspects of the modular group, the fusion algebra
and the notion of a self-dual MCFT.
The FLM Moonshine Module, $\MM$,
\ref{FLM} is historically the first example  of
such a  construction being a $\Z_2$ orbifolding of the Leech lattice
MCFT, $\VL$.  We  review the
usefulness of these ideas in understanding Monstrous
Moonshine, the genus zero property for Thompson series \ref{CN}
which we have shown is equivalent
to the property that the only meromorphic $\Z_n$ orbifoldings
of $\MM$ are $\VL$ and  $\MM$ itself (assuming that $\MM$
is uniquely determined by its characteristic function
$J(\tau)$) \ref{T1,T2}. We show that these
constraints on the possible $\Z_n$ orbifoldings of $\MM$
 are also sufficient to demonstrate the genus zero property
for Generalised Moonshine functions in the
simplest non-trivial prime cases by considering $\Z_p\times\Z_p$
orbifoldings of $\MM$. Thus Monstrous Moonshine implies
Generalised Moonshine in these cases.

\bigskip
{\bf 1. Meromorphic Conformal Field Theory.}
In this section, we will review some of the basic properties
of Meromorphic Conformal Field Theory (MCFT) (or chiral algebras) as
described by Goddard \ref{Go}. This is a physically motivated
approach to Vertex Operator Algebras \ref{B1,FLM,FHL}
containing the same essential ideas.
Let $\H$ denote some Hilbert space with a dense subspace of states
$\{\phi\}$ including a
unique \lq vacuum state\rq\  $\vac$ with properties described below.
 In a MCFT we define a set of conformal fields or vertex operators
$\V$ such that corresponding to each state
$\phi $ there exists an \op $V(\phi,z)\in \V$ acting on $\H$  with
$$
\eqalign{
\lim_{z\rightarrow 0} V(\phi,z) \vac = \phi}
\eqno (1.1)
$$
It is assumed that there exists Virasoro \ops $L_n$ which form the modes
of $V(\om,z)$ (see below) for a Virasoro state $\om$ where
$$
\eqalignno{
[L_n,V(\phi,z)]=&z^n[z{d\over dz}+(n+1)h_\phi]V(\phi,z)&(1.2a)\cr
[L_m,L_n]=&(m-n)L_{m+n}+{C\over 12}m(m^2-1)\delta_{m,-n}&(1.2b)\cr
}
$$
where $h_\phi$ is called the conformal weight of $\phi$ and
$C$ is the central charge for the representation of the Virasoro algebra
(1.2b).
The Virasoro state $\om$ has conformal weight 2.
{}From (1.2a), $L_0$ defines a discrete grading on $\H$ with
 $L_0 \vert \phi\rangle =h_\phi \vert \phi \rangle$. We assume that $\V$
 is unitary so that $h_\phi\ge 0$.
By a {\it Meromorphic CFT}, we will mean a CFT for
which the conformal weights are integral
and where the \ops $\V$ obey the (bosonic) Locality Property :
$$
\eqalign{
V(\phi,z) V(\psi,w)=V(\psi,w) V(\phi,z)
}
\eqno (1.3)
$$
with  $\vert z\vert>\vert w\vert $ on the LHS analytically continued to
$\vert z\vert<\vert w\vert $  on the RHS.
(These assumptions  ensure
that all correlation functions are meromorphic).
These operators can  then be shown to
satisfy the Duality Property \ref{Go,FHL}
$$
\eqalign{
V(\phi,z)V(\psi,w)=V(V(\phi,z-w)\psi,w)
}
\eqno(1.4)
$$
with $\vert z\vert >\vert w\vert $ and $\vert z-w\vert <\vert w\vert$
respectively and
where $V(\phi,z)$ is extended by linearity to any state in $\H$. These
are  essentially the defining properties of a vertex (operator) algebra as
defined in \ref{B1} and developed in \ref{FLM}.
All the conformal fields in a MCFT also
obey the Monodromy condition :
$$
\eqalign{
V(\phi,e^{2\pi i}z)=V(\phi,z)
}
\eqno (1.5)
$$
so that  the mode expansion for each operator is
$V(\phi,z)= \sum_{k\in \Z} \phi_k z^{-k-h_\phi}$ with
$\phi_k\vac=0$ for all $k >-h_\phi$ and $\phi_{-h_\phi}\vac=\phi$
from (1.1) e.g. the modes for the Virasoro (energy-momentum)
\op $V(\om,z)$ are $\{L_n\}$ as above with $\om=L_{-2}\vac$.
Then (1.4) leads to an exact form of
the usual operator product expansion of CFT \ref{BPZ}
$$
\eqalign{
V(\phi,z)V(\psi,w)=\sum_{k=0}^\infty (z-w)^{k-h_\phi-h_\psi}V(\chi,w)
}
\eqno(1.6)
$$
where $\chi=\phi_{-k+h_\psi}(\psi)$ is a state of conformal weight $k$.
We will sometimes schematically write such an expansion as $\V \V\sim \V$.

\bigskip

{\bf 2. The Modular Group and Self-Dual MCFTs.}
Let $\V$ be a MCFT and define the characteristic function
(or partition function) for $\V$ by the following trace
$$
\eqalign{
Z(\tau)={\roman Tr}_{\H}(q^{L_0-C/24})
}
\eqno (2.1)
$$
where $\tau \in H$, the upper half complex plane.
In string theory models, $Z(\tau)$ arises when finding the probability
for a closed string to form a 2-torus parameterised by $\tau$.
The simplest example, is the one-dimensional  $C=1/24$ bosonic string which
has characteristic function $1/\eta(\tau)$ where
$\eta(\tau)=q^{1/24}\prod_{n>0} (1-q^n)$. For a $d$ dimensional
$C=d/24$ string compactified by an even lattice $\L$, we obtain
a MCFT denoted by $\VL$, with
$Z(\tau)=\Theta_\L(\tau)/[\eta(\tau)]^d$  where
$\Theta_\L=\sum_{\lambda\in\L} q^{\lambda^2/2}$.

The behaviour of $Z(\tau)$ under the action of the modular
group $\Ga={\roman SL}(2,\Z)$, generated by $T:\tau\rightarrow \tau+1$ and
$S:\tau\rightarrow-1/\tau$,
is  related to the meromorphic properties of $\V$ and to properties of
the meromorphic irreducible representations of $\V$.
For a MCFT we clearly have
$Z(\tau+1)=e^{2\pi iC/24}Z(\tau)$ and, in particular, $Z(\tau)$ is
$T$ invariant for $C=24$.

Let us now discuss the meaning of the $S$ transformation.
Let $\tilde\V$ denote an irreducible meromorphic representation
for $\V$ acting on a Hilbert space $\H^K$
and let $\V^K$ be the corresponding set of intertwiners acting on $\H$ that
create the states of $\H^K$ from the original vacuum vector
$\vac\in \H$ \ref{FHL,DGM}. Then as in (1.3) and (1.4) we have
$$
\eqalignno{
\tilde V(\phi,z) \tilde V(\psi,w)=
&\tilde V(\psi,w)\tilde V(\phi,z)&(2.2a)\cr
\tilde V(\phi,z) \tilde V(\psi,w)=
&\tilde V(V(\phi,z-w)\psi,w)&(2.2b)\cr
\tilde V(\phi,z) V^K(\chi,w)=
&V^K(\chi,w) V(\phi,z)&(2.2c)\cr
\tilde V(\phi,z) V^K(\chi,w) =
&V^K(\tilde V(\phi,z-w)\chi,w)&(2.2d)\cr
}
$$
(up to suitable analytic continuations) for $\tilde V(\phi,z)\in\tilde \V$,
$V^K(\chi,z)\in \V^K$ with  $\phi,\psi\in\H$ and $\chi\in\H^K$.
Given such a representation, we thus naturally extend $\V$ to act
on $\H\oplus\H^K$ and henceforth we drop the tilde notation
distinguishing the space on which $\V$ acts.
We also define the characteristic function $Z^K(\tau)=\Tr{\H^K}{}$
for $\V^K$. In general, the conformal weights of $\H^K$ are not
integral but are equal $\mod {\Z}$ and hence $Z^K(\tau)$ is $T$
invariant up to a phase.

By a {\it Rational MCFT}, we will mean a MCFT
 which has a finite number $M$ of such irreducible representations
$\{\V^K\}$, $K=0,...,M-1$ (with $\V\equiv \V^0$)
 and where every representation of $\V$ is reducible.
For a Rational MCFT, Zhu has shown that
each characteristic function $Z^K(\tau)$ is
holomorphic on the upper half plane $H$ (given a certain
growth condition which is conjectured to follow from
rationality) and the functions $\{ Z^K\}$
transform amongst themselves under the modular group $\Ga$ \ref{Z}.

These properties can also be understood  if $\V$
together with (possibly multiple copies of) its
intertwiners form a non-meromorphic
CFT which we call the {\it Dual CFT} to $\V$ and
denote by $\Vst$. We can think of $\Vst$ as comprising
the maximal (in some sense!) set of vertex
\ops of central charge $C$ that are local with respect to $\V$.
$\Vst$ is expected to satisfy an  operator product algebra given by
some generalised version of (1.6) where schematically
$$
\eqalign{
\V^{I} \V^{J} \sim \sum_{K=0}^{M-1} N^{IJK} \V^K}
\eqno(2.3)
$$
where $N^{IJK}$ are non-negative integers determining the
decomposition in terms of irreducible representations of $\V$
of the non-meromorphic algebra - these are
the {\it Fusion Rules} for  $\Vst$ \ref{Ve}. The coefficients $N^{IJK}$
satisfy a commutative associative algebra which is diagonalised by
$S :Z^I\rightarrow S^{IJ} Z^J$ where $S^{IJ}$ is a unitary symmetric
matrix \ref{Ve}.
In addition, we assume $\Vst$ is a unitary CFT, so that
$S^{I0}/S^{00}\ge 1$ with equality
iff we have {\it Abelian Fusion Rules} i.e.
 for every given $I,J=0,...M-1$, $N^{IJK}=1$ for some
unique $K$ so that every pair of intertwiners
fuses to form a unique intertwiner.
Assuming Abelian Fusion Rules we then find,
since $S^I_J$ is symmetric and unitary, that
$$
\eqalign{
S:Z(\tau)\rightarrow \epsilon_S{1\over \sqrt M}\sum_K Z^K(\tau)=
\epsilon_S\vert\Hst/\H\vert^{-1/2} \Tr{\Hst}{}
}
\eqno (2.4)
$$
where $\vert \epsilon_S\vert=1$ and
$\Hst$ denotes the Hilbert space $\oplus_{K=0}^{M-1} \H^K$ for
$\Vst\equiv \oplus_K \V^K$ and $M=\vert \Hst/\H\vert$.
If furthermore
$\{ Z^K\}$ is charge conjugation invariant then $S^{IJ}$ is real so that
$\epsilon_S=1$. This formula can
be verified for an even lattice $\L$ MCFT  where
the irreducible representations for  $\VL$ are indexed by $\Lst/\L$
where $\Lst$ is the dual lattice \ref{D1}.  In this case, $\VL$
is naturally embedded in the non-meromorphic CFT $\V^{\Lst}$
so that $(\VL)^\ast = \V^{\Lst}$. Furthermore,
the fusion rules are abelian from the abelian structure of $\Lst$ \ref{DL}.
Then,  under the action of
$S$, $Z_\L=\Theta_\L/\eta^d \rightarrow
\vert \Lst/\L \vert^{-1/2} \Theta_\Lst/\eta^d$
in the usual way in agreement with (2.4) with $\epsilon_S=1$. Similarly,
(2.4) holds with $\epsilon_S=1$
for the  Abelian orbifold constructions discussed below.

If $Z(\tau)$ is $S$ invariant and hence $\V$
is the unique irreducible representation
for itself,  we define $\V$ to be a {\it Self-Dual MCFT}.
This is only possible for $C=0\mod{8}$ \ref{Go}.
For $C=24$, then $Z(\tau)$ is
modular invariant with a unique simple pole at $q=0$ on $H/\Ga$
which is equivalent to the Riemann sphere of genus zero.
Hence $Z(\tau)$ is given by $J(\tau)$, the {\it hauptmodul} for $\Ga$
\ref{Se}
$$
\eqalignno{
Z(\tau)=&J(\tau)+N_0&(2.5a)\cr
J(\tau)=&{E_2^3\over{\eta^{24}}}-744
={1\over q}+0+196884q+ 214 93760 q^2 + ... &(2.5b)\cr
}
$$
with $E_2(\tau)$ the
Eisenstein modular form of weight 4 \ref{Se} and
where $N_0$ is the number of conformal weight 1 operators in $\V$.
Examples of such theories are lattice models where $\L$ is a Niemeier
even self-dual lattice. Then  $\VL$ is meromorphic self-dual because
$\L$ is even self-dual. In particular, for the Leech lattice which
contains no roots,  $N_0=24$. Other examples of self-dual C=24 MCFTs
are the Moonshine Module $\MM$ with $Z(\tau)=J(\tau)$ and other orbifold
constructions as we now describe. In general, there are thought to be just 71
such independent self-dual MCFTs \ref{Sch}.

\bigskip
{\bf 3. Abelian Orbifolding of a Self-Dual MCFT.}
Let $\V$ be a self-dual MCFT and let $Aut(\V)$
denote the \auto group preserving the operator algebra
for $\V$ with
$$
\eqalign{
gV(\phi,z)g^\m1=V(g\phi,z)
}
\eqno(3.1)
$$
for each $g\in Aut(\V)$. Consider $G$ any finite abelian subgroup of $Aut(\V)$
generated by $m$ commuting elements $\{g_1,...,g_m\}$ of order $n_1,...,n_m$.
Let $\PG\V$ denote the \ops invariant under $G$ with projection operator
$\PG= {1\over \vert G\vert }\sum_{g\in G}g$. $\PG\V$ is a MCFT
but is not self-dual as can be seen by studying the
corresponding characteristic function $\Tr{\PG \H}{}$ which is not $S$
invariant.
In particular, consider the trace for each $g\in G$
$$
\eqalign{
Z(1,g)\equiv\Tr{\H}{g }
}
\eqno(3.2)
$$
where we introduce standard notation indicating  boundary conditions
on the 2-torus where the first label $1$ refers to the monodromy
condition (1.5).  Using path integral methods in string theory \ref{DHVW}
one can argue that under $S:\tau\rightarrow -1/\tau$ the boundary conditions
are
interchanged for $Z(1,g)$ charge conjugation invariant and
$\V$ a self-dual theory so that
$$
\eqalign{
S:Z(1,g)\rightarrow Z(g,1)=\Tr{\Hg}{}=D_gq^{E_0^g}+...
}
\eqno(3.3)
$$
the characteristic function for $\Hg$, the \lq $g$-twisted\rq\ Hilbert space.
We assume that $\Hg$ is uniquely defined (up to isomorphism)
for each $g\in G$.
The parameters $E_0^g$ and  $D_g$ are called the $g$-twisted vacuum
energy and degeneracy.
Note also that the remaining coefficients of the powers are
necessarily all non-negative integers.  We assume that each
twisted  state $\psi\in\Hg$ of conformal weight $h_\psi$ is
created from $\vac$ by the action of a twisted \op
with the following $g$-twisted  Monodromy property:
$$
\eqalign{
V(\psi,e^{2\pi i}z)= gV(\psi,z)g^\m1=e^{-2\pi i h_\psi}V(\psi,z)
}
\eqno(3.4)
$$
We denote the set of such \ops for each $g\in G$ by $\Vg$.
We also assume that $\oplus_{g\in G}\Vg$ satisfies a
non-meromorphic version of the Locality property (1.3) and a
$G$-invariant operator product expansion
generalising (1.6) (up to suitable analytic continuation) where
$$
\eqalignno{
V(\psi,z) V(\chi,w)=
&\epsilon_{\psi,\chi} V(\chi,w) V(\psi,z)&(3.5a)\cr
V(\psi,z) V(\chi,w)=
&\sum_{\rho}(z-w)^{h_\rho-h_\psi-h_\chi}V(\rho,w)&(3.5b)\cr}
$$
for $\psi\in\Hg$, $\chi\in\Hh$ and $\rho\in \Hgh$ for $g,h\in G$. The Locality
phase $\epsilon_{\psi,\chi}$ is of order dividing $\vert G\vert$
and is unity for $\phi\in \PG\H$ and any $\psi\in \Hg$. This latter
property implies  that each twisted sector $\Vg$ is the intertwiner for
a meromorphic representation of $\PG\V$ as in (2.2c,d). This
representation can be then further decomposed into $\vert G\vert$
irreducible representations $\Vg=\oplus_{j_k}\Vg^{\{ j_k \}}$ labelled by
the eigenvalues $\{\exp{2\pi ij_k/n_k}\}$ of the generators $\{g_k\}$ for $G$.
Thus, in the language of the last section,
we have a set of $\vert G\vert^2$
irreducible representations for the Rational MCFT
$\PG\V$ which together form the Dual CFT
given by $(\PG\V)^\ast =\oplus_{g,j_k} \Vg^{\{ j_k \}}$ with
Abelian Fusion Rules :
$\Vg^{\{ i_k \}}\Vh^{\{ j_k \}}\sim\Vgh^{\{i_k+j_k\}}$
from (3.5b). Then (2.4) is recovered with $\epsilon_S=1$  using (3.3) where
$$
\eqalign{
S:Z(1,\PG) =\Tr{\PG\H}{}\rightarrow
{1\over \vert G\vert}\sum_{g\in G}Z(g,1)
= {1\over \vert G\vert}\Tr{(\PG\H)^\ast}{}
}
\eqno (3.6)
$$
where $(\PG\H)^\ast \equiv \oplus_{g\in G} \Hg$.

Since the \ops of $\Vg$ are
eigenvalues of $g$, the centraliser of $C(g\vert Aut(\V))$ has a natural
extension as the \auto group, which we denote by $C_g$,
 of the non-meromorphic algebra $\V\Vg\sim \Vg$. This extension
depends on the $g$-twisted vacuum degeneracy $D_g$.
Defining $G_n=C(g\vert Aut(\V))/\gen{g}$,
in general one finds that $C_g=\hat L.G_n$ for some
extension $\hat L=\gen{g}.L$ determined by
the \auto group acting on the twisted vacuum of $\Hg$.
(Here $A.B$ denotes a group with normal subgroup $A$ where $B=A.B/A$).
If the twisted
vacuum is unique ($D_g=1$), then $C_g=\gen{g}\times G_n$. For each $h\in C_g$
we can then generalise (3.2) to define
$$
\eqalignno{
&Z(g,h)=\Tr{\Hg}{h}&(3.7a)\cr
&T: Z(g,h)\rightarrow Z(g,g^\m1 h),\quad S: Z(g,h)\rightarrow
Z(h,g^\m1)&(3.7b)\cr
}
$$
which transform under $T$ as given in
(3.4) and under $S$ by an interchange of $g$ and $h$ boundary conditions
assuming (3.3) in general.  Then for any $\gamma=
(\matrix{ a&b\cr c&d \cr})
$ in $\Ga$,
$\gamma : Z(g,h)\rightarrow Z(h^{-c} g^{a},h^{d} g^{-b})$.
In particular, for all $g,h\in G$, these characters form a basis for
the characters of the irreducible representations
$\Vg^{\{ j_k \}}$ of the Rational MCFT $\PG\V$ \ref{DVVV}.
Thus, each $Z(g,h)$ is expected to be holomorphic on $H$ \ref{Z}.
Other important properties of $Z(g,h)$ are that
given charge conjugation invariance
then $Z(g,h)=Z(g^\m1,h^\m1)$ so that $\gamma$ and $-\gamma$ act equally
for each $\gamma\in \Ga$.  Finally,
given the uniqueness of the twisted
sectors, it also clear that under conjugation by any element $x\in Aut(\V)$,
with $g\rightarrow g^x=xgx^\m1$, then $x(\Vg) x^\m1$ is isomorphic to
$\V_{g^x}$ so that $Z(g,h)=Z(g^x,h^x)$ for all $x\in Aut(\V)$.

The construction of \ops  obeying (3.4)
is only known in string theory-like
models \ref{DHVW,L,DGM,DM1} where
the \auto $g$ is lifted  from an \auto of the embedding space
of the string, typically a lattice automorphism.
The properties of (3.5) are assumed in the physics
literature \ref{DFMS,DVVV} and are only so far understood in
limited settings for vertex operator algebras \ref{H}. The modular
transformation properties (3.7b) for $Z(g,h)$ can be explicitly demonstrated in
many cases \ref{Va,DM1}.

The $G$ orbifold MCFT is now constructed from the projection
$\VorbG=\PG((\PG\V)^\ast)$ which has
characteristic function $Z_{\roman orb}=\sum_{g,h\in G}Z(g,h)$.
In general, $g$ may act projectively on $\Vg$ in (3.4)
for a given $g\in G$ of order $n$ so that $g^n$ is a global phase.
Then $\VorbG$ is not meromorphic and $Z_{\roman orb}$ is
not $T$ invariant. Such a \lq global phase anomaly\rq\ is absent whenever
$nE_0^g=0\mod{1}$ \ref{Va} so that the \ops of $\Pg\Vg$ are of integer
conformal weight. Assuming no such anomalies arise then $\VorbG$
is a self-dual MCFT and so $Z_{\roman orb}(\tau)=
J(\tau)+N_0^{\roman orb}$ as in (2.5) where
$N_0^{\roman orb}$ is the number of conformal weight 1 \ops in $\VorbG$.

The OPA (3.5) is also preserved by the action of the
dual \auto group $G^\ast$, defined as follows. Recalling that
$G=\gen{g_1,...,g_m}$ with $g_k$ of order $n_k$, we
define $g_k^\ast$  by
$$
\eqalign{
g_k^\ast V(\psi,z) g_k^{\ast-1} = e^{2\pi ir_k/n_k}V(\psi,z)
}
\eqno(3.8)
$$
for each $\psi\in \Vg$ where $g=g_1^{r_1}...g_m^{r_m}$.  Then
 $G^\ast =\gen{g_1^\ast,... ,g_m^\ast}$ is clearly an \auto group
for (3.5) and is isomorphic to $G$.
We may then consider the orbifolding of $\VorbG$ with
respect to $G^\ast$. The $G^\ast$ invariant \ops of $\VorbG$ are
$\P_{G^\ast}\VorbG =\PG\V$ as before. Therefore the
projection of the dual is $P_{G^\ast}(\PG\V)^\ast=\V$ i.e.
orbifolding $\VorbG$ with respect to $G^\ast$ reproduces $\V$.
 Thus the two self-dual MCFTs $\V$ and $\VorbG$ are placed on an
 equal footing with each  an Abelian orbifolding of the other.
 Thus we have :
$$
\eqalign{
\matrix{
 (\PG\V)^\ast  \cr
{\buildrel \P_{G^\ast}\ \ \over\swarrow} \quad
 {\buildrel\ \  \PG\over\searrow}  \cr
 \V\quad
         \matrix{{\buildrel G\over \longrightarrow}  \cr
                {\buildrel G^\ast\over \longleftarrow}}
           \quad\VorbG\cr}\cr
{\buildrel \ \ \PG\ \ \over\searrow} \quad
 {\buildrel \P_{G^\ast}\  \over\swarrow}\quad  \cr
\PG\V \qquad\ \cr}
\eqno (3.9)
$$
where the horizontal arrows represent an orbifolding with respect
to the indicated  \auto group and the diagonal arrows are projections.

\bigskip

{\bf 4. The Moonshine Module and Monstrous Moonshine.}
The FLM Moonshine module  $\MM$
is historically the first example of a self-dual orbifold MCFT \ref{FLM}
and is constructed as a $\Z_2$ orbifolding of $\VL$, which will denote the
Leech lattice MCFT from now on.  The $\Z_2$ \auto $r$
of $\VL$ chosen is lifted from the
lattice reflection $\r$ so that $\Pr\VL$ contains no conformal weight
1 operators. The $r$-twisted space ${\H}_r$ on the other hand has
vacuum energy $E_0^r=1/2$ (and is hence global  phase anomaly free)
but likewise contains no conformal weight 1 operators since $E_0^r>0$.
The resulting orbifold MCFT, $\MM=\Pr(\VL\oplus\Vr)$, therefore has
characteristic function $J(\tau)$. As shown by FLM,
a symmetrisation of the vertex algebra of the
196884 conformal weight 2 \ops (including the Virasoro \op
$V(\om,z)$) forms an affine version of the 196883
dimensional Griess algebra \ref{Gr} whose \auto group is the Monster $\M$.
FLM went on to show that $\M=Aut(\MM)$ \ref{FLM}.
Note that we can identify as in (3.8),
the \auto group $\gen{r^\ast}$ dual to $\gen{r}$.
By considering $Aut(\Pr\VL)$ and
$Aut(\Pr\Vr)$, the centraliser $C(r^\ast\vert \M)$ can be found to be
$C(r^\ast\vert \M)= 2^{1+24}_{+}.\Co$ where $\Co$ denotes the Conway simple
group (i.e. the \auto group $\roman Co_0$ of $\L$ modulo $\r$),
$2^{1+24}_{+}$ is an extra-special 2-group.
Then $\M$ is generated by $C$ and another
involution that mixes the twisted and untwisted sectors \ref{Gr,FLM}.
Furthermore, $\MM$ can be orbifolded with respect to $\gen {r^\ast}$ as in
(3.9)
to recover $\VL$.

 FLM have conjectured that $\MM$ is characterised (up to isomorphism) as
follows \ref{FLM}:

\proclaim $\MM$ Uniqueness Conjecture.
$\MM$ is the unique CFT with characteristic function $J(\tau)$.

\noindent This is stated in the context of the assumptions of
Sections 1 and 2  where $Z(\tau)=J(\tau)$ is modular invariant
and hence $\MM$ is a self-dual C=24 MCFT. Furthermore, $\MM$ forms
the unique irreducible representation for itself \ref{D2}.
We will now consider briefly some evidence for this conjecture.

We may consider other possible $\Z_n$ orbifoldings of $\VL$  with
characteristic function $J(\tau)$ which should reproduce $\MM$
according to this  conjecture.  In general, we can classify all
\autos $a$ of $\VL$ lifted from \autos $\a\in {\roman Co}_0$,
(for which $\Va$ can be explicitly constructed)
so that \ref{T2}

\item{(i)} $\Pa\VL$ contains no conformal dimension 1 \ops i.e.
$\a$ is fixed point free.
\item{(ii)} $E_0^a>0$ i.e. $\Va$ contains no conformal dimension 1 operators.

\item{(iii)} $\Va$ is global phase anomaly free i.e. $nE_0^a=0\mod{1}$ for
$\a$ of order $n$.

There are 51 classes of ${\roman Co}_0$
obeying (i) and (ii) only and 38 classes satisfying
(i),(ii) and (iii). These 38 classes include 5 prime ordered cases
for which $(p-1)\vert 24$. These have been considered in much greater detail by
Dong and Mason \ref{DM2} who reconstructed $\MM$ exactly for $p=3$ and
by Montague who also analysed the $p=3$ case \ref{M}. For each of these
38 classes, we expect that a self-dual MCFT $\Vorba$
with characteristic function
$J(\tau)$ exists. Furthermore, orbifolding $\Vorba$
with respect to the dual group $\gen{a^\ast}$
defined as in (3.8) reproduces  $\VL$ with $\V=\VL$, $G=\gen{a}$
and $\VorbG=\MM$ in (3.9).
By analysing $Aut(\Pa\V_{a^k})$ for
$k=0,...,n-1$  we can calculate explicitly the centraliser \ref{T2}
$$
\eqalign{
C(a^\ast\vert Aut(\Vorba))=&\hat L.G_n
}
\eqno(4.1)
$$
where $G_n=C(\a\vert{\roman Co}_0)/\gen{\a}$ and $\hat L=n.L$ is
a cyclic extension of $L=\L/(1-\a)\L$.
For the prime ordered cases, this reduces to a well-known centraliser
formula for $\M$ \ref{CN}. (4.1) can also be shown
to hold for all 51 classes obeying (i) and (ii) once $a^\ast$ is
appropriately defined and is verified for $Aut(\Vorba)=\M$ in many
cases \ref{T2}. All of this provides evidence that $\Vorba=\MM$
in each construction lending weight to the uniqueness conjecture.
Further evidence is given below.

Let us now define the Thompson-McKay series $\Tgt$ for each $g\in \M$
$$
\eqalign{
\Tgt=\Tr{\HMM}{g}={1\over q}+0+[1+\chi_A(g)]q+...
}
\eqno (4.2)
$$
where $\chi_A(g)$ is the character of the 196883 dimensional adjoint
representation for $\M$.  This trace is obviously reminscent of (3.2)
and this interpretation will be further explored below.
The Thompson series for the identity element is $J(\tau)$, which
is the hauptmodul for the genus zero modular group
$\Ga={\roman SL}(2,\Z)$ as already stated.
By calculating the first ten terms of $\Tgt$ for each conjugacy class of $\M$,
Conway and Norton \ref{CN} conjectured

\proclaim Monstrous Moonshine.
For each $g\in\M$, $\Tgt$ is the hauptmodul for a genus zero fixing
modular group $\Gg$.

\noindent Borcherds has now demonstrated this rigorously
although the origin of the genus zero property remains obscure \ref{B2}.
In general, for $g$ of order $n$,
$\Tgt$ is found to be invariant under
$\Go{n}=\{(\matrix{a& b\cr
              nc & d\cr})\vert {\roman det }=1\}$ up to $h^{\roman th}$ roots
of unity
where $h\vert n$ and $h\vert 24$.
$\Tgt$ is fixed by $\Gg$ with $\Go{N}\subseteq \Gg$ and contained in
the normaliser of $\Go{N}$ in ${\roman SL}(2,{\R})$ where $N=nh$ \ref{CN}.
This normaliser always contains the Fricke involution
$W_N:\tau\rightarrow -1/N\tau$ where $W_N^2=1 \mod{\Go{N}}$.
We will refer to those classes with $h=1$ as {\it Normal}
and those with $h\not =1$ as {\it Anomalous} i.e.
the fixing group of $\Tgt$ is of type $n+e_1,e_2,...$  for normal
classes and of type $n\vert h+e_1,e_2,...$ for anomalous classes
in the notation of \ref{CN}. This  terminology is motivated by whether
the corresponding twisted sector $\Vg$ described
below has a global phase anomaly or not.

For a normal element $g\in \M$ of prime order $p$ (there is only
one anomalous prime class of order 3 with $h=3$) we find
either $\Gg=\Go{p}$ or $\Go{p}+=\gen{\Go{p},W_p}$.
$\Go{p}$ is of genus zero only when
$(p-1)\vert 24$. There is a corresponding class of $\M$, denoted
by $p-$, for each such prime with this Thompson series e.g. the involution
$r^\ast$ above belongs to the class $2-$.
$\Go{p}+$ is of genus zero for all the prime
divisors of the order of $\M$.
There is a class of $\M$, denoted by $p+$, for each such prime with
Thompson series fixed by $\Go{p}+$.  In general all the classes of $\M$ can be
divided into Fricke and non-Fricke classes according to whether or not
$\Tgt$ is invariant under the Fricke involution $W_N$.
It is also observed that the Thompson series for Fricke classes
have non-negative integer coefficients whereas the
coefficients of non-Fricke Thompson series are integers
of mixed sign.  There are a total of 51 non-Fricke classes of which 38  are
normal
and there are a total of 120 Fricke classes of which 82 are normal.

For each of the 38 constructions above
based on classes $\{\a\}$ satisfying the conditions (i)-(iii)
we can compute the dual \auto Thompson series $T_{a^\ast}$ and this
agrees precisely with the genus zero series for the 38
non-Fricke normal classes of the Monster which also obey the centraliser
relationship (4.1). Likewise, we can identify the other 13 anomalous
non-Fricke classes and find the corresponding correct genus zero
Thompson series \ref{T2}.  This is further evidence for the
assertion that $\Vorba=\MM$ implied by the uniqueness conjecture for $\MM$
which we will now assume to be true from now on.

We now turn to the interpretation of a Thompson series
as an orbifold trace with $\Tgt=Z(1,g)$ as in (3.2) where now $\V=\MM$.
For $g$ in a non-Fricke class, we can construct the twisted sector $\Vg$
by choosing $g=a^\ast$ as above with characteristic function obeying
(3.3).  In particular, all the coefficients of $Z(g,1)$ are non-negative
integers and hence $Z(1,g)-T_g(0)$, which is inverted up to a multiplicative
constant under the Fricke involution to give $Z(g,1)(N\tau)-T_g(0)$,
has mixed sign coefficients
as observed. For the 38 normal non-Fricke classes
we may orbifold $\MM$ with respect to
$\gen{a^\ast}$ to obtain $\VL$.
Then the vacuum energy $E_0^g=0$ for the twisted sector
$\Vg$ so that conformal weight 1 \ops are reintroduced.
On the other hand, for an anomalous non-Fricke class,
a global phase anomaly parameterised by the
parameter $h\not =1$  occurs and  we cannot obtain a MCFT
from the resulting orbifolding \ref{T1}.

Consider next $f\in \M$ a Fricke element of order $n$.
The corresponding twisted sector
can be constructed when $f$ is lifted from a lattice automorphism.
We will assume that $\Vf$ exists in each case obeying (3.3)-(3.5).
For normal elements, no global phase anomaly occurs and we may orbifold
$\MM$ with respect to $\gen{f}$ to obtain a self-dual MCFT
$\Vorbf$.  Assuming $\Tft$ is a hauptmodul we then find that
$\Vorbf=\MM$  for each normal Fricke element. The converse
is also true, where given that $\Vorbf=\MM$ for some $f\in \M$ then $T_f$
is the hauptmodul for a genus zero modular group containing the Fricke
involution \ref{T2}. In general, we find (assuming the uniqueness
conjecture for $\MM$) that for all
normal elements of $\M$
$$
\eqalign{
 \VL \
\matrix{{\buildrel \gen{a}\over \longrightarrow}  \cr
            {\buildrel \gen{a^\ast}\over \longleftarrow}}\
\MM\   {\buildrel \gen{f}\over \longleftrightarrow}\ \MM
}
\ \Leftrightarrow \ \hbox{$T_{a^\ast}$, $T_f$ are hauptmoduls}
\eqno (4.3)
$$

(4.3) can be understood briefly for the prime ordered
normal Fricke classes as follows. Suppose that $f$ is a $p+$ element with
Fricke invariant hautpmodul $\Tft$. Then
$Z(f,1)(\tau)=Z(1,f)(\tau/p)=q^{-1/p}+0+...$ so that $\Vf$ has vacuum energy
$E_0^f=-1/p$, degeneracy $D_f=1$, contains no conformal weight 1 operators
and has non-negative integer coefficients. Thus
$\Pf\Vf$ does not reintroduce conformal weight 1 operators.
Similarly $\Pf\V_{f^k}$, $k\not =0\mod{p}$,
contains no such operators ($f$ and $f^k$ are conjugate) so that
$\Vorbf=\MM$ since the characteristic function is $J(\tau)$.

Conversely, if $\Vorbf=\MM$ for a prime $p$ ordered element $f$ then
since $f$ and $f^k$ are conjugate,
$\Tft$ is automatically $\Go{p}$ invariant.
The fundamental region $H/\Go{p}$ for $\Go{p}$ has only two cusp
points \ref{Gu} at $\tau=\infty$ where $\Tft$ has a simple pole of order 1
from (2.2) and at $\tau=0$ which is singular iff $E^f_0< 0$ with
residue $D_f$ from (3.3). We can then argue that since $\Vorbf=\MM$,
$E_0^f=-1/p$ with $D_f=1$. This follows by considering
the dual \auto $f^\ast\in\M$ to $f$ as in (3.9)
and showing that $T_{f\ast}=T_f$. Then the corresponding centralisers
must be equal which implies that $D_f=1$, since no extension occurs.
Furthermore, $\Vf$ contains no conformal weight 1 \ops which
implies that either $E_0^f=-1/p$ or $E_0^f>0$. The latter possibility
is ruled out because then $\Tft=q^\m1+0+O(q)$ would be a hauptmodul for
$\Go{p}$
which implies $E_0^f=0$ when the constant term of $T_f$ is zero.
Thus we must have $E_0^f=-1/p$ with $D_f=1$.
Finally, consider $\phi(\tau)=\Tft-T_f(W_p(\tau))$ which
is $\Go{p}$ invariant and is holomorphic on
the compactification of $H/\Go{p}$, which is a compact Riemann surface.
Hence $\phi(\tau)$ is a constant which
is zero since it is odd under $W_p$. Hence $\Tft$ is $\Go{p}+$
invariant and has a unique simple pole on $H/\Go{p}+$ and is
therefore a hauptmodul for $\Go{p}+$.

This argument can be generalised
to any normal Fricke element $f\in \M$ of order $n$.
Then (4.3) is equivalant to the fact that
(i) $\Vf$ has vacuum energy $E_0^f=-1/n$
and degeneracy $D_f=1$ and  (ii) if $f^r$ is Fricke then so is
$f^s$ with $s=n/(r,n)$ where $r$ and $s$ must be co-prime.
These conditions are then sufficient to supply all the poles and
residues of $\Tft$ so that
$\Tft$ is a hauptmodul for some genus zero fixing group \ref{T1,T2}.
Finally, the genus zero property for an anomolous class of $\M$,
which follows from the Harmonic formula of \ref{CN}, is described in \ref{T2}.

\bigskip

{\bf 5. Generalised Moonshine from Abelian Orbifolds.}
Let us now consider the more general set of conjectures suggested by Norton
\ref{N} concerning Moonshine for centralisers (or extensions thereof) of
elements of the Monster $\M$.  Specifically, in the notation of (3.7a)  we
consider:
$$
\eqalign{
Z(g,h)= \Tr{\Hg}{h}
}
\eqno (5.1)
$$
for $h\in C_g\equiv {Aut}(\Vg)$. For all Fricke elements, the
twisted Hilbert space vacuum,
which we now denote by $\Hg\vert_0$, is unique and hence $C_g=
\gen{g} \times G_n$ where $G_n=C(g\vert \M)/\gen{g}$ whereas for $g$ non-Fricke
$C_g=\hat K.G_n$ (for some extension $\hat K$).
Norton has conjectured :

\proclaim Generalised Moonshine Conjecture.
$Z(g,h)$ is either constant or is a hauptmodul for some genus zero
fixing group for every pair of commuting elements $g,h\in \M$.


\noindent This conjecture has been explicitly verified for an orbifold
construction based on the Mathieu group $M_{24}$ \ref{DM1}.
In terms of the orbifold picture reviewed in the earlier
sections we can note the following properties for $Z(g,h)$:

(i) $Z(g,h)=Z(g^{-1},h^{-1})$.

(ii)  $Z(g,h)=Z(g^x,h^x)$  for conjugation with respect to any $x\in \M$.

(iii) $S: T_g(\tau)\rightarrow Z(g,1)=D_g q^{E_0^g}+...$ is a series with
non-negative integer coefficients decomposible into positive sums of the
 dimensions of the  irreducible representations of $C_g$ where
 $E_0^g$ is twisted vacuum energy and $D_g$ is the vacuum degeneracy, the
dimension of $\Hg\vert_0$, the twisted Hilbert space vacuum.

(iv) From (3.7) we find that
$\gamma : Z(g,h)\rightarrow Z(h^{-c} g^{a},h^{d} g^{-b})$
 for $\gamma=
(\matrix{ a&b\cr c&d \cr})
\in \Ga$.
Note that $g=\exp(-2\pi i E_0^g)$ on
$\Hg\vert_0$. In particular, for a normal Fricke  element of order $m$,
$g=\om=e^{2\pi i/m}$ and each $h\in C_g$ acts as some element of
$\gen{\om}$ on $\Hg\vert_0$.

(v) As a consequence of (iv), $Z(g,h)$ is invariant up to roots of unity
under $\Gmn{m}{n}=
\{ \gamma\in \Ga\vert a=1\mod{m},b=0\mod{m},c=0\mod{n},d=1\mod{n}\} $
where $m=O(g)$, $n=O(h)$. These extra factors appear if $h^{-c} g^a$ is
anomalous for some co-prime $a$ and $c$.

(vi) The value of $Z(g,h)$ at any parabolic cusp $a/c$ ($a$ and $c$ co-prime)
is determined by the vacuum energy of the $k=g^a h^{-c}$
twisted sector from (iv).
In particular, only the Fricke classes are responsible for singular
cusp points \ref{N}. The residue of these cusps is determined by the
action of $h^d g^{-b}$ on $\H_k\vert_0$. We will assume,
as discussed in section 3, that
$Z(g,h)$ is holomorphic at all other points on $H$.

Thus given any commuting pair of elements $g,h$ as above, the location
of any singularities for $Z(g,h)$ is known by finding which of the classes
$k=g^a h^{-c}$ is Fricke for $(a,c)=1$. The strength of the pole is then
determined by the corresponding vacuum energy $E_0^k$. However, the residue
for each singular cusp still needs to be found. We will argue below that
this extra information is also supplied by the constraints of (4.3),
at least in the simplest non-trivial prime
cases.  Once these singularities are known,
then $Z(g,h)$ can be shown in each case to be either constant or
to be the hauptmodul for a genus zero modular group.

The basic idea is to consider the orbifolding of $\MM$
 with respect to $\gen{g,h}$ and to re-express this as the composition of two
$\Z_n$ orbifoldings. If $\gen{g,h}=\Z_k$,  $k= mn/(m,n)$, then $Z(g,h)$ can
always  be related to a regular Thompson series via an appropriate modular
transformation e.g. for $m, n$ co-prime  with
$am+bn=1$ then $\gen{g,h}=\Z_{mn}$ and $
(\matrix{ 1& 1\cr -bn&am \cr})
: Z(1,gh)\rightarrow Z(g,h)$ from (iv).
For $\gen{g,h}\not =\Z_k$ we will consider here
the simplest non-trivial case where
$\gen{g,h}$ contains only normal prime order $p$ elements.
Then $Z(g,h)$ is $\Gn{p}\equiv\Gmn{p}{p}$ invariant from (v). We will
further assume that $h\con{C_g} h^k$ for $k\not = 0\mod{p}$   i.e. conjugate
in $C_g$. This is sufficient to ensure that the coefficients in the
$q$ expansion of $Z(g,h)$ are rational
since all the irreducible characters  are rational.
Furthermore, this condition restricts
the possible conjugacy  classes in $\M$  generated by $g$ and $h$ to just
three i.e. $g\con{\M} g^a$, $h\con{\M} h^b$ and $gh\con{\M} g^a h^b$
for $a,b \not= 0\mod{p}$.
{}From (iv), we have that $Z(g,h)$ is fixed by $\Goo{p}=
\{\gamma\in \Ga\vert b=c=0\mod{p}\}\sim \Go{p^2}$ (under conjugation by
${\roman diag}(1,p)$ so that $Z(g,h)(p\tau)$ is $\Go{p^2}$ invariant).
$Z(g,h)$ therefore has parabolic cusps on $H/\Goo{p}$
at $\tau=i\infty, 0,1,...,p-1$ \ref{Gu} with behaviour determined, from (vi),
by the vacuum energy of the sectors twisted
by $g, h,g^{p-1}h,...g^2h, gh$ respectively where the last $p-1$ classes are
conjugate.   Within these assumptions we then
find that there are 5 possible cases (up to relabelling) that
may occur for any $p$ as follows.

{\bf Case 1 : }  $g, h, gh= p-$.
 None of the cusps are singular
and therefore $Z(g,h)$ is holomorphic on $H/\Goo{p}$ and hence is constant.

We may now assume for the remaining 4 cases (without loss of
generality by relabelling) that $g=p+$  so that
$$
\eqalign{
Z(g,h)=q^{-1/p}+0+O(q^{1/p})}
\eqno(5.2)
$$
We also note from (iv)  that $g$ acts as $\om=e^{2\pi i/p}$ on $\Hg\vert_0$.

{\bf Case 2 : } $g=p+, \ h,  gh=p-$.
 In this case $Z(g,h)$ has a unique simple pole
 at $q=0$ as in (5.2) on  $H/\Goo{p}$ and therefore $Z(g,h)$ is a hauptmodul.
 This is only possible for $p=2,\ 3,\ 5$ (where $Z(g,h)(p\tau)$ is a hauptmodul
 for $\Go{p^2}$). For $p=5$, no such Generalised Moonshine function is actually
 observed which, interestingly, is also the case for regular Monstrous
Moonshine
 where $25-$ is one of the so-called ghost elements \ref{CN}.

{\bf Case 3 :} $ g, h=p+,\ gh=p-$.
 $Z(g,h)$ has two singularities at
$\tau=i\infty$ and $0$. Under $S:\tau\rightarrow-1/\tau$ we have
$$
\eqalign{
Z(g,h) \rightarrow Z(h,g^\m1)=\om^{-k_g}q^{-1/p}+0+O(q^{1/p})
}
\eqno (5.3)
$$
where $g=\om^{k_g}$ on $\Hg\vert_0$, $k_g\in\Z_p$ .
We  may conjugate $h$ to $h^\m1$ in $C_g$
so that $Z(h,g^\m1)=Z(h^\m1,g^\m1)=Z(h,g)$, from (i),
which implies that $2k_g=0\mod{2}$. Hence for $p>2$, $k_g=0\mod{p}$. For $p=2$
we will show below that $k_g=0\mod{2}$ also. Consider $f(\tau)=
Z(g,h)(\tau)-Z(g,h)(S(\tau))=0+O(q^{1/p})$. $f(\tau)$ is
$\Goo{p}$ invariant without any poles on $H/\Goo{p}$ and hence is constant and
equal to zero. Therefore, $Z(g,h)$ is $\gen{\Goo{p},S}\sim \Go{p^2}+$ invariant
with a unique simple pole and is a hauptmodul. This is only possible for
$p=2,\ 3,\ 5,\ 7$. For $p=7$, no such Generalised Moonshine function is
observed which  corresponds to the ghost element $49+$ of Monstrous Moonshine !

To understand the $p=2$ case it is necessary to consider the interpretation
of $Z(g,h)$ in terms of a $\gen{g,h}=\Z_2\times\Z_2$ orbifolding of $\MM$.
The orbifold so obtained is meromorphic self-dual (since no anomalous Monster
elements occur) and is explicitly
$$
\eqalign{
\Vorbgh= \Pgh(\MM\oplus\Vg\oplus\Vh\oplus\Vgh)}
\eqno(5.4)
$$
where
$\Pgh=(1+g+h+gh)/4=\Pg\Ph$. We can consider this as two successive $\Z_2$
orbifoldings
$$
\eqalign{
\Vorbgh=\Pg(\Ph(\MM\oplus\Vh)\oplus\Ph(\Vg\oplus\Vgh))}
\eqno(5.5)
$$
i.e. $\Vorbgh$ is a $Z_2$ orbifolding with respect to $\gen{g}$
of $\Vorbh\equiv \Ph(\MM\oplus\Vh)$.
Since $h=2+$, we know that $\Vorbh =\MM$ and hence
$\Ph(\Vg\oplus\Vgh)$ is a $g$ twisted sector for $\MM$ for $g$ of order two
by the assumed uniqueness of the twisted sectors. Thus
$g=2+$ or $ 2-$ when acting on $\Vorbh$. However,
we can determine from (5.5) that the character for this $g$ twisted sector is
$[Z(g,1)+Z(g,h)+Z(gh,1)+Z(gh,h)]/2=q^{-1/2}+...$  using (5.2).
This implies that $g$ is Fricke when acting on $\Vorbh$ and hence
$\Vorbgh=\MM$ with $g=2+$. We can represent this sequence of orbifoldings
diagramatically as follows:
$$
\matrix{
 \ &   \ & \MM & \ & \ \cr
         \ &   {\buildrel \gen{h}\ \over \nearrow}
         &\   & {\buildrel \ \gen{g} \over \searrow} & \ \cr
         \MM& \ & {\buildrel \gen{g,h}\over \longrightarrow} & \ & \MM \cr
}
\eqno (5.6)
$$
where each copy of $\MM$  is orbifolded with respect to the denoted group.

We can similarly consider the orbifolding of $\MM$
with respect to $g=2+$ followed by $h$. The resulting
orbifold must be $\Vorbgh=\MM$ and hence $\Pg(\Vh\oplus\Vgh)$
must be a 2+ twisted sector. This forces $g=1$ on $\Hh\vert_0$ as was claimed
 earlier.  In general, for any $p$, it is straightforward to see that
$\MM{\buildrel \gen{g}\over \longrightarrow}\Vorbg=
\MM {\buildrel \gen{h}\over \longrightarrow}  \Vorbgh=\MM$ in this case.

{\bf Case 4 :} $g, gh=p+, \ h=p-$.
In this case $Z(g,h)$ has singular cusps at
$i\infty, 1,...,p-1$ on $H/\Goo{p}$.   We can find the residues of these
poles by decomposing the orbifolding
with respect to $\gen{g,h}$ to obtain
$\MM{\buildrel \gen{g}\over \longrightarrow}
\MM {\buildrel \gen{h}\over \longrightarrow}  \Vorbgh$  since $g=p+$  where we
necessarily  find that either $\Vorbgh=\MM$ or $\VL$ from (4.3).
If we alternatively orbifold with respect to $h=p-$ first we then obtain
$\MM{\buildrel \gen{h}\over \longrightarrow}
\VL {\buildrel \gen{g}\over \longrightarrow}  \Vorbgh$.
In order that $\Vorbgh=\MM$, it is necessary that the $g$ twisted sector
of $\VL$ so obtained, $\Ph(\oplus_{k=0}^{p-1}\V_{gh^k})$, has positive
 vacuum energy  from condition (ii) of section 4.
However, from (5.2), this is impossible
since $\Ph\Vg$ has character  $q^{-1/p}+0+O(q^{1/p})$. Hence $\Vorbgh=\VL$ in
this case.

We can similarly decompose $\Pgh=\P_{gh}\Pf$ for $f=g^a h$ a $p+$ element
with $a\not =0,1 \mod{p}$.
Then $\MM{\buildrel \gen{f}\over \longrightarrow}
\MM {\buildrel \gen{gh}\over \longrightarrow}  \Vorbgh=\VL$. This implies
that $gh$ must act as a $p-$ element on $\Vorb{f} =\MM$ and hence the
corresponding $gh$ twisted sector $\P_{f}(\oplus_{k=0}^{p-1}\V_{gh f^k})$
has zero vacuum energy from (4.3).
In particular, this implies that $\P_{f}\Hgh\vert_0=0$
so that  $f=g^a h\not =1$ on $\Hgh\vert_0$ for any $a\not =0\mod{p}$
(noting that $gh=\om$ on $\Hgh\vert_0$). Let $h=\om^r$ be the action on
$\Hgh\vert_0$ (from (iv)) so that $g=\om^{1-r}$.
But we can always choose $a\not =0 \mod{p}$ such that
$g^a h $ acts as unity on $\Hgh\vert_0$
unless $r= 0 \mod {p}$. Hence the orbifolding is only consistent when $h=1$ on
$\Hgh\vert_0$. In general, by conjugation, we then find that
$h=1$ on $\H_{g^a h^b}\vert_0$ for all $b\not= 0\mod{p}$.  Hence, the residue
of any of the singular cusps is known.
This allows us to find the full fixing modular group.

Let $\gamma_p=
(\matrix{
1& -p\cr
1 & 1-p\cr})$
which is of order $p$ in
$\Goo{p}$. $\gamma_p$ permutes the $p$ cusps of $Z(g,h)$ where
$\gamma_p:Z(g,h)\rightarrow Z(gh, h^\m1)=q^{-1/p}+0+O(q^{1/p})$
and similarly for the other singular twisted sectors.
Then $f(\tau)=Z(g,h)(\tau)-Z(g,h)(\gamma_p(\tau))$ is
holomorphic on $H/\Gn{p}$ and is therefore zero. Hence $Z(g,h)$ is
$\gen{\Goo{p},\gamma_p}\sim \Go{p}$ invariant with   a unique simple pole
and is a hauptmodul. This is only possible for $p=2,\ 3,\ 5,\ 7,\ 13$.
Once again, the largest possible case is not observed, $p=13$, although this
does not correspond to a ghost element for regular Moonshine.

{\bf Case 5 :} $g ,h, gh=p+$.
In this last case all sectors are Fricke and
$Z(g,h)$ has singular cusps at $i\infty, 0,1,...,p-1$. With the assumption
that all $h\con{C_g}h^k$ for $k\not=0 \mod{p}$ we need only in practice
consider $p=2,3$ and 5
where $C_g=\gen{g}\times G_p$ for $G_p=B, Fi'_{24}$ or $HN$ respectively
\ref{CCNPW}. Following  a general
argument as in Case 3, it is easy to see again that $\Vorbgh=\MM$
since both $g$ and $h$ are Fricke and (5.2) is obeyed.

For  $p=2$ we again  decompose the orbifolding of
$\MM$ with respect to $\gen{g,h}$.
Referring to (5.5), we note that $\Ph(\Vg\oplus\Vgh)$
has a $2+$ character and that $gh$ is Fricke. Hence $h=-1$ and
$g=1$ on $\Hgh\vert_0$. Similarly, we can  orbifold with respect to $g$ first
and find that $\Pg(\Vh\oplus\Vgh)$ also has $2+$ character. Hence $g=-1$
on  $\Hh\vert_0$. Thus all the residues of $Z(g,h)$ are known. In particular,
$ST$ of order three permutes the cusps $\{i\infty,0,1\}$ with
$ST:Z(g,h)\rightarrow Z(gh,g)=q^{-1/2}+0+O(q^{-1/2})$. Then, by the usual
argument, $Z(g,h)$ is a hauptmodul for $\gen{\Ga(2),ST}$ of genus zero,
which is of level 2 and index 2 in $\Ga$.
In fact,  $Z(g,h)$ is invariant under the full modular group $\Gamma$ up
to $\pm 1$ with $Z(g,h)=-Z(g,gh)$ so that
$Z(g,h)(2\tau)=E_2(\tau)/\eta^{12}(\tau)-252$ is the
hauptmodul for $2\vert 2$ in the notation of \ref{CN}.

For $p=3,5$ we may repeat the argument of Case 3 to
show that $g=1$ on $\Hh\vert_0$ so that $Z(h,g)=q^{-1/p}+0+...O(q^{1/p})$.
But $Z(g,h)$ and $Z(h,g)$ have
the same singular structure and hence we may
interchange $g$ and $h$. Since $\Vgh$ is preserved by this interchange,
$g=h$ on $\Hgh\vert_0$ with $gh=\om$ so that $g=h=\om^{(p+1)/2}$. Hence
by conjugation as in (ii), all the residues of the
singular cusps of $Z(g,h)$ are known.

For $p=3$, let  $\gamma_2=T^\m1 S T$ which is of order 2 and interchanges
the cusps $\{\infty,0\}\leftrightarrow\{2,1\}$ whereas $S$
interchanges $\{\infty,1\}\leftrightarrow\{0,2\}$. Then
$\gamma_2:Z(g,h)\rightarrow Z(gh,h^\m1g)=q^{-1/3}+0+...$
and $S:Z(g,h)\rightarrow
Z(h,g^\m1)=q^{-1/3}+0+...$ and similarly for the other cusps. By the usual
argument, we then find that $Z(g,h)$ is the hauptmodul
for the genus zero group $\gen{\Ga(3),S,T^\m1ST}$ of level 3 and index 3 in
$\Ga$. Further analysis shows that $Z(g,h)$ is invariant under $\Ga$ up to
third roots of unity with $Z(g,h)=\om^2 Z(g,gh)=\om Z(g,g^2h)$,
so that $Z(g,h)(3\tau)=E_3(\tau)/\eta^{8}(\tau)-368$ is the
hauptmodul for $3\vert 3$ in the notation of \ref{CN}.

For $p=5$, let $\gamma_3=T S T^3$  which is of order 3 and cyclically
permutes the  cusps $\{\infty,0\} \rightarrow\{1,4\}\rightarrow\{2,3\}$
whereas $S$ interchanges the cusps $\{\infty,1,2\}\leftrightarrow\{0,4,3\}$.
Then $\gamma_3:Z(g,h)\rightarrow Z(gh^\m1,g^3h^3)=
Z(g h, (gh^\m1)^3)= q^{-1/5}+0+...$ by conjugation and
similarly for the other cusps. $Z(g,h)$ is invariant under $\Ga(5)$ whose
normaliser contains $\Ga$ and so $Z(g,h)(\tau)-Z(g,h)(\gamma(\tau))$
is holomorphic on $H/\Ga(5)$ for both $\gamma=S$ and $\gamma_3$ and hence is
zero. Thus $Z(g,h)$ is the hauptmodul for the
genus zero group $\gen{\Goo{5},S,\gamma_3}$ which is of level 5 and
index 5 in $\Ga$. In the notation of \ref{FMN}, the fixing group of
$Z(g,h)(5\tau)$ is $5\vert\vert 5$. In this case, there are five independent
functions $f_k(\tau)=Z(g,g^k h), k=0,1,...,4$ which are permuted under $\Ga$.

We summarise Cases 2-5 in the following table where we reproduce the
genus zero fixing group for $Z(g,h)(p\tau)$ with $g=p+$ and
where only the actual observed values of $p$ are indicated.

\medskip

\vbox{\tabskip=0pt \offinterlineskip
\def\tabrule{\noalign{\hrule}}

\vskip 10 pt\halign{
\vrule\  \hfil$#$\hfil \  & \vrule \  \hfil$#$\hfil \  &
\vrule \  \hfil$#$\hfil \vrule\cr
\tabrule
\ & gh=p- & gh=p+ \cr
\tabrule
h=p- &\  \Go{p^2}, p=2,3\  &\ \Go{p}-, p=2,3,5,7\  \cr
\tabrule
h=p+ &\  \Go{p^2}+, p=2,3,5\ &\  2\vert 2, 3\vert 3, 5\vert\vert 5\ \cr
\tabrule
}}

\bigskip

{\bf 6. Conclusion.}
We have shown that the genus zero property for the Generalised Moonshine
functions (5.1) follows from the genus zero property for Thompson
series in the simplest non-trivial prime cases. It remains a much greater
challenge to extend this arguments to all cases.
The major difficulties of this method for general commuting
elements $g,h$ are (i) the proliferation
of possible Fricke classes in $\gen{g,h}$ giving the location
of poles and (ii) the determination of the corresponding residues.
Once this information is known, then any generalised moonshine functions
should be reconstructible if it is a hauptmodul.
Finally, it is interesting that the ghost groups $25-$ and $49+$ are
absent from the above table (as indeed is $50+50$ from the list
of modular groups for the centraliser moonshine of
the $5+$ or $10+$ elements of $\M$ where it might be expected to arise).
These hauptmoduls are also distinguished by having non-quadratic
irrationalities at their non-singular cusps \ref{FMN} suggesting some
possibly deeper number theoretic significance.

\bigskip

\centerline{\bf References.}

\medskip\item{\ref{B1}}
Borcherds, R.,
{\it Vertex algebras, Kac-Moody algebras and the monster},
\hfill\break Proc.Natl.Acad.Sc.USA {\bf 83} (1986) 3068-3071.

\medskip\item{\ref{B2}}
Borcherds, R.E.,
{\it Monstrous moonshine and monstrous Lie superalgebras},
\hfill\break Invent.Math.{\bf 109} (1992) 405-444.

\medskip\item{\ref{BPZ}}
Belavin,A.A.,  Polyakov, A.M.  and  Zamolodchikov, A.B.,
{\it Infinite conformal symmetry and two dimensional quantum field theory},
Nucl.Phys. {\bf B241} (1984) 333-380.

\medskip\item{\ref{CCNPW}}
Conway, J.H., Curtis, R.T.,  Norton, S.P.,  Parker, R.A. and  Wilson, R.A.,
An atlas of finite groups,
(Clarendon Press, Oxford, 1985).

\medskip\item{\ref{CN}}
Conway, J.H. and  Norton, S.P.,
{\it Monstrous Moonshine},
Bull.London.Math.Soc. {\bf 11} (1979) 308-339.

\medskip\item{\ref{DFMS}}
Dixon, L.,  Friedan, D.,  Martinec, E. and Shenker, S.,
{\it The conformal field theory of orbifolds},
Nucl.Phys. {\bf B282} (1987) 13-73.

\medskip\item{\ref{DGM}}
Dolan, L.,  Goddard, P.  and Montague, P.,
{\it Conformal field theory of twisted vertex operators},
Nucl.Phys. {\bf B338} (1990) 529.

\medskip\item{\ref{DHVW}}
Dixon, L.,  Harvey, J.A.,  Vafa, C.  and Witten, E.,
{\it Strings on orbifolds},
Nucl.Phys. {\bf B261} (1985) 678;
{\it Strings on orbifolds II},
Nucl.Phys. {\bf B274} (1986) 285-314.

\medskip\item{\ref{D1}}
Dong, C.,
{\it Vertex algebras associated with even lattices},
J.Algebra {\bf 161} (1993) 245-265.

\medskip\item{\ref{D2}}
Dong, C.,
{\it Representations of the moonshine vertex operator module},
Univ.Cal.Santa Cruz preprint (1992).

\medskip\item{\ref{DL}}
Dong, C and Lepowsky, J., {\it Abelian intertwiner algebras - a generalization
of vertex operator algebras}, preprint 1993.

\medskip\item{\ref{DM1}}
Dong,C. and Mason, G.,
{\it An orbifold theory of genus zero associated to the sporadic group
$M_{24}$}   Commun.Math.Phys. {\bf 164} (1994) 87-104.

\medskip\item{\ref{DM2}}
Dong, C.  and Mason, G.,
{\it On the construction of the moonshine module as a ${\Z}_p$ orbifold},
U.C.Santa Cruz Preprint 1992.

\medskip\item{\ref{DVVV}}
Dijkgraaf, R.,  Vafa, C.,  Verlinde, E. and Verlinde, H.,
{\it The operator algebra of orbifold models},
Commun.Math.Phys. {\bf 123} (1989) 485-526.

\medskip\item{\ref{FHL}}
Frenkel, I., Huang, Y.-Z.  and Lepowsky, J,
{\it On axiomatic approaches to vertex operator algebras and modules},
Mem.A.M.S. {\bf 1-4}, 1993.

\medskip\item{\ref{FLM}}
Frenkel, I.,  Lepowsky, J.  and Meurman, A.,
{\it A natural representation of the Fischer-Griess monster
with the modular function  $J$ as character},
 Proc.Natl.Acad.Sci.USA {\bf 81} (1984) 3256-3260;
{\it A moonshine module for the monster} in:
J.Lepowsky et al. (eds.), Vertex operators in mathematics and physics,
(Springer Verlag, New York, 1985);
Vertex operator algebras and the monster,
(Academic Press, New York, 1988).

\medskip\item{\ref{FMN}}
Ford, D., McKay, J. and Norton, S.P.,
{\it More on replicable functions}, preprint, 1993.

\medskip\item{\ref{Go}}
Goddard, P.,
{\it Meromorphic conformal field theory} in:
Proceedings of the CIRM Luminy conference, 1988, Page 556,
(World Scientific, Singapore, 1989).

\medskip\item{\ref{Gr}}
Griess, R.,
{\it The friendly giant},
Inv.Math. {\bf 68} (1982) 1-102.

\medskip\item{\ref{Gu}}
Gunning, R.C.,
Lectures on modular forms,
(Princeton University Press, Princeton, 1962).

\medskip\item{\ref{H}}
Huang, Y.-Z.,
{\it A non-meromorphic extension of the moonshine
vertex operator algebra}, Rutgers Univ. preprint 1994; and these Proceedings.

\medskip\item{\ref{L}}
Lepowsky, J.,
{\it Calculus of twisted vertex operators}, Proc.Natl.Acad.Sci.USA
{\bf 82} (1985) 8295-8299.

\medskip\item{\ref{M}}
Montague, P.,
{\it Codes, lattices and conformal field theories},
Cambridge University Ph.D.  dissertation, 1991.

\medskip\item{\ref{N}}
Norton, S. P.,
{\it Generalised moonshine},
Proc.Symp.Pure Math. {\bf 47} (1987) 209-210.

\medskip\item{\ref{Se}}
Serre, J-P.,
A course in arithmetic,
(Springer Verlag, New York, 1970).

\medskip\item{\ref{Sch}}
Schellekens, A.N.,
{\it Meromorphic C=24 conformal field theories}
Commun.Math.Phys. {\bf 153} (1993) 159-185

\medskip\item{\ref{T1}}
Tuite, M.P.,
{\it Monstrous  moonshine from orbifolds},
Commun.Math.Phys. {\bf 146} (1992) 277-309.

\medskip\item{\ref{T2}}
Tuite, M.P.,
{\it On the relationship betwen monstrous moonshine
 and the uniqueness of the moonshine module},
DIAS-STP-93-09, To appear,  Commun.Math.Phys.

\medskip\item{\ref{Va}}
Vafa, C.,
{\it Modular invariance and discrete torsion on orbifolds},
Nucl.Phys. {\bf B273} (1986) 592-606.

\medskip\item{\ref{Ve}}
Verlinde, E.,
{\it Fusion rules and modular transformations
in 2-d conformal field theory},
Nucl.Phys. {\bf B300} (1988) 360-376.

\medskip\item{\ref{Z}}
Zhu, Y.,
{\it Vertex operator algebras, elliptic functions and modular forms},
Yale Ph.D. dissertation, 1990; and these Proceedings.

\end